%% file: icsea2026_paper.tex
\documentclass[conference]{IEEEtran}
\IEEEoverridecommandlockouts

\usepackage{amsmath,amssymb,amsfonts}
\usepackage{graphicx}
\usepackage{booktabs}
\usepackage{multirow}
\usepackage{xcolor}
\usepackage{tikz}
\usetikzlibrary{shapes.geometric, arrows.meta, positioning, fit, backgrounds, calc}
\usepackage{soul}
\sethlcolor{yellow}
\soulregister{\itshape}{0}
\usepackage{pgfplots}
\pgfplotsset{compat=1.18}
\usepackage[hidelinks]{hyperref}
\usepackage{url}
\usepackage{orcidlink}
\usepackage{float}
\usepackage{footmisc}

\begin{document}

\title{\fontsize{14}{16.8}\selectfont\bfseries Trustworthy RAG: An Evaluation Agent for Detecting Misinformation and Knowledge Poisoning in Generative AI Systems}

\author{
\IEEEauthorblockN{
Balkrishna Giri\,\orcidlink{0009-0006-0576-7392},
Md Toufique Hasan\,\orcidlink{0009-0005-8907-9898},
Jussi Rasku\,\orcidlink{0000-0002-4401-8013},
Muhammad Waseem\,\orcidlink{0000-0001-7488-2577},
and Pekka Abrahamsson\,\orcidlink{0000-0002-4360-2226}
}
\IEEEauthorblockA{
Faculty of Information Technology and Communication Sciences, Tampere University\\
Tampere, Finland\\
\resizebox{0.95\textwidth}{!}{{\ttfamily e-mail: \{balkrishna.giri, mdtoufique.hasan, jussi.rasku, muhammad.waseem, pekka.abrahamsson\}@tuni.fi}}\\[0.2em]
\parbox{0.95\textwidth}{
\centering
\scriptsize\bfseries
Published in the Proceedings of the Twenty-First International Conference on Software Engineering Advances (ICSEA 2026), IARIA, 2026.
}
}
}

\IEEEaftertitletext{\vspace{-1\baselineskip}}
\maketitle

\begin{abstract}
Retrieval-Augmented Generation (RAG) grounds Large Language Model (LLM) outputs in external knowledge, but RAG systems usually trust whatever they retrieve, creating a \emph{Security-Reliability Gap}: high semantic relevance does not guarantee factual truth. Adversaries exploit this through \emph{knowledge poisoning}, inserting malicious documents to cause targeted misinformation. We propose an \emph{Evaluation Agent}, middleware that combines Natural Language Inference (NLI) factual verification, a five-signal poison detector with relevance-weighted aggregation, and a \emph{Trust Index} $T = 0.4\,F + 0.35\,C + 0.25\,(1-P)$ with a non-linear dampener for high-contamination contexts. On TruthfulQA with Llama~3.3~70B, the agent reaches 91\% accuracy and 100\% precision, with 100\% recall on instruction injection, while in-place edits, such as entity swaps, remain hard to detect. Across three LLMs the Trust Index stays discriminative, with a Receiver Operating Characteristic Area Under the Curve (ROC-AUC) of 0.73 to 0.81; generation style matters more than model size, and per-LLM threshold calibration restores baseline-competitive accuracy, whereas a weaker FEVER result shows that cross-dataset generalization requires domain-specific calibration. In a software-engineering use case, a secure-coding assistant over guidance from the Open Worldwide Application Security Project (OWASP) Top~10 and the Common Weakness Enumeration (CWE), the agent reliably blocks instruction injection of unsafe advice (F1 92\%), while contradiction and subtle semantic weakening remain hard. Throughout, the agent measures \emph{detection} of poisoned context before generation, not whether the LLM adopts the injected misinformation. We release the proposed approach, attack generator, and experimental artifacts at the link: \url{https://github.com/GPT-Laboratory/TrustworthyRAG}.
\end{abstract}


\input{sections/01_introduction.tex}
\input{sections/02_background.tex}

\input{sections/03_approach.tex}
\input{sections/04_experimental_design.tex}
\input{sections/05_results.tex}

\input{sections/06_discussion.tex}

\input{sections/07_conclusion.tex}

\section*{Author Contributions}
Balkrishna Giri and Md Toufique Hasan contributed equally to this work.

\section*{Acknowledgment}
This work was partly supported by the AI Native Software Engineering (ANSE) project, funded by Business Finland. The authors declare no conflicts of interest.

\input{sections/08_references.tex}

\end{document}

%% file: sections/01_introduction.tex
\section{Introduction}

Large Language Models (LLMs) built on the Transformer architecture \cite{vaswani2017attention} now power knowledge-intensive applications, such as question answering, search, and code assistance, yet their parametric knowledge is frozen at training time and they often hallucinate, producing fluent text that is factually wrong \cite{ji2023survey}. Retrieval-Augmented Generation (RAG) \cite{lewis2020rag} addresses this by retrieving documents from an external corpus and conditioning generation on them, and it has become a default pattern for deploying LLMs over private or fast-changing knowledge \cite{gao2024retrieval}.

Grounding generation in retrieved text creates a new dependency: the answer is only as trustworthy as the corpus behind it. Standard RAG assumes a benign knowledge base, and existing evaluation frameworks, such as RAGAS \cite{es2024ragas} and ARES \cite{saadfalcon2024ares}, measure faithfulness to the retrieved context rather than the integrity of that context. In open-world deployments, however, an adversary can insert malicious documents through knowledge poisoning \cite{zou2024poisoned}; injecting only five malicious passages per question into a corpus of millions ($\approx$0.0002\%) can drive attack success to roughly 90\%. A system that answers faithfully from a poisoned document therefore scores well under current metrics while emitting compromised output. We call this the \emph{Security-Reliability Gap}: high semantic relevance is treated as a proxy for truth, and the few available poisoning defenses are mostly offline corpus-cleaning steps with no online trust layer inside the inference loop.

This paper presents an \emph{Evaluation Agent} that closes this Security-Reliability Gap. It acts as defensive middleware that screens retrieved context before generation and outputs an interpretable \emph{Trust Index} that fuses Natural Language Inference (NLI) factual verification \cite{maynez2020faithfulness}\cite{honovich2022true}, a five-signal poison detector (Section~\ref{sec:approach}), and a cross-document consistency estimate. We evaluate the agent on two public benchmarks (TruthfulQA and FEVER) across three LLMs and four attack strategies, and then apply it to a software-engineering use case, a secure-coding assistant that retrieves from guidance based on the Open Worldwide Application Security Project (OWASP) Top~10 and the Common Weakness Enumeration (CWE). We study three research questions:
\begin{itemize}
  \item \textbf{RQ1.} How effectively does the Evaluation Agent detect misinformation and knowledge poisoning in retrieved context?
  \item \textbf{RQ2.} How does NLI-based factual verification change the trustworthiness of RAG output?
  \item \textbf{RQ3.} How resilient is the approach across LLMs, datasets, and attack strategies, including a secure-coding setting in the software-development lifecycle (SDLC)?
\end{itemize}

The RQs operationalize this gap: RQ1 targets detection, RQ2 the NLI trust layer in the inference loop, and RQ3 robustness across generators, domains, and attacks; Section~\ref{sec:discussion} answers each explicitly. Code, prompts, the attack generator, and experimental artifacts are publicly available \cite{repoartifacts}.

We find that the agent detects overt poisoning well (100\% recall on instruction injection, 91\% accuracy and 100\% precision on TruthfulQA mixed attacks) but that in-place edits, such as entity swaps and subtle weakening, stay near-undetectable, which is a limit of surface-signal detection rather than of tuning. We also show that NLI-based trust scoring depends on the generating LLM (generation style matters more than model size), is stable across repeated runs, and is invariant to retrieval depth, while a weaker FEVER result bounds external validity. These findings extend prior RAG evaluation by adding an online check on context integrity, showing where such checks succeed and where world-knowledge verification is needed.


%% file: sections/02_background.tex
\section{Background and Related Work}
\label{sec:background}

\textbf{RAG architectures.} RAG has developed from retrieval-augmented language models \cite{guu2020realm} and retrieve-then-generate pipelines \cite{izacard2021fid} into modular systems \cite{gao2024retrieval}. Modular RAG separates the retriever, generator, and evaluator into exchangeable components. Self-RAG \cite{asai2024selfrag} adds reflection tokens for retrieval and critique, but it is unclear whether self-correction can resist adversarial inputs designed to manipulate the critique process.

\textbf{Evaluation.} Retrieval quality is often measured with Information Retrieval (IR) metrics, such as top-$k$ retrieval accuracy \cite{karpukhin2020dense} and Mean Reciprocal Rank (MRR). These metrics measure relevance, but not safety. A poisoned document that is highly relevant to a query can still score well. Generation quality has moved from n-gram overlap toward factuality-based measures that separate faithfulness from factual correctness \cite{maynez2020faithfulness}. TruthfulQA \cite{lin2022truthfulqa} focuses on imitative falsehoods, and the TRUE benchmark \cite{honovich2022true} supports factual-consistency evaluation with metrics, such as NLI. Recent hallucination detectors mainly verify generated text rather than the integrity of the corpus. SelfCheckGPT \cite{manakul2023selfcheckgpt} measures self-consistency across sampled generations, while fine-grained factuality scores \cite{min2023factscore} check atomic facts against a trusted knowledge source. RAGAS \cite{es2024ragas} and ARES \cite{saadfalcon2024ares} use LLMs to judge faithfulness and relevance, but they do not evaluate whether the retrieved corpus itself is trustworthy.

\textbf{Attacks and defenses.} Indirect prompt injection places malicious instructions inside retrievable content \cite{greshake2023indirect}, and poisoning the web-scale datasets used for model pre-training can be done at low cost \cite{carlini2024poisoning}. PoisonedRAG \cite{zou2024poisoned} studies knowledge poisoning in the inference corpus, and adversarial passages can be optimized so that they enter the retrieved set for many queries \cite{zhong2023poisoning}. On the defense side, retrieval augmentation can reduce hallucination \cite{shuster2021retrieval}, and NLI-based verification can compare a retrieved document as the premise with the generated answer as the hypothesis to estimate factual consistency \cite{maynez2020faithfulness}\cite{honovich2022true}. However, many poisoning defenses are still \emph{offline}, such as filtering or cleaning the corpus before indexing. These methods cannot detect poisoned content that appears at inference time. Toxicity classifiers also miss misinformation that is written in neutral and professional language. Closest to our setting are online RAG defenses that filter or vote over the retrieved passages themselves: certifiably robust aggregation generates an answer per isolated passage and securely aggregates the results \cite{xiang2024robustrag}, and trust-aware retrieval filtering discards suspicious passages before generation \cite{cheng2025raguard}. These defenses act on the retriever side and often need multiple generations per query or corpus-level assumptions; our agent instead scores the retrieved set and the generated answer jointly and emits an interpretable verdict without modifying retriever or generator. Our work combines factual verification and poison detection into one interpretable online score, and shows where this type of surface-signal detection works and where it fails.

\textbf{Security in the software-development lifecycle.} Software-engineering research increasingly studies how generative Artificial Intelligence (AI) is used in development \cite{tuomisto2025ethical} and how AI is adopted in quality assurance \cite{karhu2025barriers}, and how digital experimentation supports sustainability in software-intensive industry \cite{jantti2025digital}. LLM coding assistants are now widely used during development, but they can generate insecure code, and developers may over-trust their suggestions \cite{pearce2022}\cite{perry2023}. The OWASP Top 10 for LLM Applications lists data and model poisoning and misinformation as major risks for these systems \cite{owaspllm2025}. When a coding assistant uses RAG over an organization's secure-coding guidance, such as the OWASP Top 10 \cite{owasp2021} and CWE \cite{mitrecwe}, that guidance corpus becomes a poisoning target. A single corrupted rule can lead the assistant to give unsafe recommendations. This motivates evaluating the Evaluation Agent in an SDLC setting (Section~\ref{sec:usecase}).

%% file: sections/03_approach.tex
\section{Proposed Approach}
\label{sec:approach}

\textbf{Threat model.} We assume a black-box knowledge-injection adversary that can insert or modify documents in the retrieval corpus but cannot access the model weights, user prompt, or the Evaluation Agent. The adversary's goal is to make the generator emit attacker-chosen misinformation or unsafe recommendations. This reflects RAG deployments that ingest third-party, web-sourced, or community-contributed content.

\begin{figure}[!ht]
\centering
\resizebox{0.63\columnwidth}{!}{%
\begin{tikzpicture}[
box/.style={rectangle, draw=black!70, rounded corners=3pt,
minimum width=2.6cm, minimum height=0.62cm,
text centered, align=center, font=\scriptsize},
mod/.style={rectangle, draw=black!70, rounded corners=3pt, fill=blue!10,
minimum width=2.6cm, minimum height=0.62cm,
text centered, align=center, font=\scriptsize},
outbox/.style={rectangle, draw=black!70, rounded corners=3pt, fill=green!12,
minimum width=4.6cm, minimum height=0.62cm,
text centered, align=center, font=\scriptsize},
stage/.style={rectangle, draw=black!45, dashed, rounded corners=4pt, inner sep=4pt},
arr/.style={-{Stealth}, thick, black!60}
]
\node[box, fill=black!8, minimum width=6.4cm] (inp) at (0,0)
{Input: query, retrieved docs, scores, embeddings};

\node[font=\tiny\itshape] (s1lab) at (0,-0.98) {Stage 1: pre-generation screening};
\node[mod] (pois) at (-1.6,-1.6) {Poison Detector};
\node[mod] (cons) at (1.6,-1.6) {Consistency Estimator};

\node[box, fill=black!8, minimum width=3.4cm] (gen) at (0,-2.75) {LLM generation (answer)};

\node[font=\tiny\itshape] (s2lab) at (0,-3.55) {Stage 2: post-generation verification};
\node[mod, minimum width=3.4cm] (nli) at (0,-4.15) {NLI Verifier (docs vs.\ answer)};

\node[box, fill=orange!15, minimum width=2.9cm] (trust) at (0,-5.5)
{Trust Index Calculator};
\node[outbox] (result) at (0,-6.5)
{EvaluationResult: score, level, warnings};

\begin{scope}[on background layer]
\node[stage, fit=(s1lab)(pois)(cons)] (s1) {};
\node[stage, fit=(s2lab)(nli)] (s2) {};
\end{scope}

\draw[arr] (inp.south) -- (s1.north);
\draw[arr] (s1.south) -- (gen.north);
\draw[arr] (gen.south) -- (s2.north);
\draw[arr] (s2.south) -- (trust.north);
\draw[arr] (s1.west) -- ++(-0.55,0) |- (trust.west);
\draw[arr] (trust.south) -- (result.north);

\end{tikzpicture}}
\caption{Two-stage Evaluation Agent pipeline: poison detection and consistency screening before generation (Stage~1), NLI verification of the generated answer after it (Stage~2). Color coding: gray input and generation steps, blue analysis modules, orange score fusion, green output verdict.}
\label{fig:arch}
\end{figure}
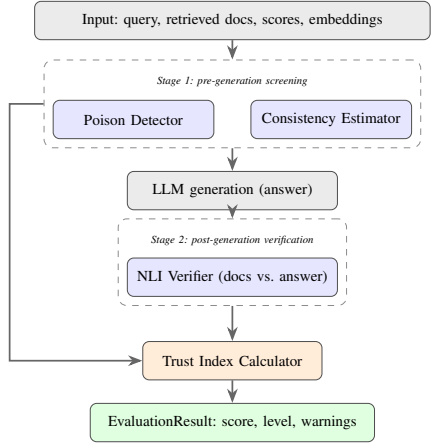

The Evaluation Agent is defensive middleware positioned between retrieval and the final trust decision (Figure~\ref{fig:arch}). It orchestrates three modules: an NLI verifier, a poison detector, and a Trust Index calculator, and emits a numeric trust score, a categorical trust level, and human-readable warnings. Operationally, the agent uses two stages. The poison detector and consistency estimator screen retrieved documents before generation, while the NLI verifier runs afterward using each document as the premise and the generated answer as the hypothesis. The signals are fused into a verdict, while deployments may also use the pre-generation signals alone.

\subsection{Factual Verification via NLI}
The NLI verifier uses \texttt{facebook/bart-large-mnli} as a sequence classifier. For each retrieved document (premise) and the generated answer (hypothesis) it produces entailment, neutral, and contradiction probabilities. The factuality score $S_{\text{factuality}}$ aggregates entailment over documents with a meaningful entailment signal; when none exists, an inconclusive baseline of $0.5$ is returned. This entailment-focused design is motivated by an empirical observation: \texttt{bart-large-mnli} assigns near-maximal contradiction ($\approx0.99$) to \emph{both} genuine contradictions and merely unrelated text, so using raw contradiction as a negative signal would systematically penalize off-topic but benign documents. Strong negative signals are therefore reserved for the dedicated poison-detection pathway.

\subsection{Multi-Signal Poison Detection}
The poison detector combines five independent signals per document: (1)~\emph{linguistic patterns} (instruction-override phrases, contradiction markers); (2)~\emph{structural anomalies} (excessive uppercase, suspicious repetition); (3)~\emph{intra-document consistency} (NLI between the first and second halves of a document); (4)~\emph{cross-document consistency} (pairwise contradiction checks, scoped so a single poisoned document cannot inflate the poison probability of clean neighbors); and (5)~\emph{semantic-outlier analysis} (embedding deviation from the batch centroid). When retrieval scores are available, document-level probabilities are combined with relevance weighting,
\begin{equation}
P_{\text{overall}} = \sum_{i=1}^{k} w_i\,P_i, \qquad w_i = \frac{s_i}{\sum_j s_j},
\end{equation}
where $s_i$ is the retrieval similarity of document $i$. This prevents low-relevance suspicious neighbors from dominating the batch verdict. The linguistic and intra-document signals target overt injections and explicit contradictions; the cross-document and semantic-outlier signals target inconsistent or anomalous insertions; and the structural signal flags formatting artifacts. The set was derived from the attack surface of the threat model, and each signal is grounded in prior observations: linguistic override patterns follow the payload style of indirect prompt injection \cite{greshake2023indirect}; the intra- and cross-document checks instantiate NLI-based consistency evaluation \cite{maynez2020faithfulness}\cite{honovich2022true}; and semantic-outlier analysis adapts embedding-space anomaly detection to the retrieved batch, motivated by corpus-poisoning attacks that insert semantically deviant passages \cite{zhong2023poisoning}. In-place value substitutions that preserve surface form fall, by construction, outside this signal set, a limit we quantify in Section~\ref{sec:usecase}.

\subsection{The Trust Index}
The Trust Index $T\in[0,1]$ is a weighted combination of three components:
\begin{equation}
\label{eq:trust}
T = \alpha\,S_{\text{factuality}} + \beta\,S_{\text{consistency}} + \gamma\,(1 - P_{\text{poison}}),
\end{equation}
with default weights $\alpha{=}0.40$, $\beta{=}0.35$, $\gamma{=}0.25$ ($\alpha{+}\beta{+}\gamma{=}1$). With these defaults, Eq.~(2) is exactly the Trust Index stated in the abstract, $T = 0.4\,F + 0.35\,C + 0.25\,(1-P)$. The weights reflect signal reliability: NLI entailment (factuality) is the most precise, cross-document agreement (consistency) a strong corpus-integrity proxy, and heuristic poison signals are weighted lowest to limit false-positive influence. Section~\ref{sec:ablation} quantifies each component's contribution and shows that the operating point is stable under moderate perturbations of these manually chosen weights.

\textbf{Non-linear dampener.} A purely linear $T$ fails under high contamination: if the LLM ignores a poisoned passage and answers from the legitimate part, factuality and consistency stay high while $P_{\text{poison}}$ is large. Because $\gamma{=}0.25$, the poison term can reduce $T$ by at most $0.25$, so $T$ can exceed the threshold $\tau{=}0.5$ despite $P_{\text{poison}} {=}0.9$. We therefore apply a multiplicative dampener when $P_{\text{poison}}>0.70$:
\begin{equation}
\label{eq:damp}
d(P) =
\begin{cases}
1.0 & P \le 0.70\\[2pt]
1.0 - 0.4\,\dfrac{P-0.70}{0.30} & P > 0.70
\end{cases}
\quad T_{\text{final}} = T\cdot d(P).
\end{equation}
The dampener is continuous at the threshold ($d(0.70){=}1$), bounded ($d(1.0){=}0.6$, i.e.\ at most $-40\%$), and proportional to contamination confidence. For the masking example ($T{=}0.59$, $P{=}0.9$), $d(0.9){=}0.733$ gives $T_{\text{final}}{=}0.43<\tau$, correctly flagging the response. For clean contexts the dampener is inactive; the multiplier falls smoothly with contamination, so only high-confidence poisoning is penalized. A secondary modifier reduces $T$ when the best retrieval similarity falls below $0.30$ (low-confidence retrieval). The final score maps to four trust levels (HIGH/MEDIUM/LOW/VERY~LOW); binary decision uses $\tau{=}0.5$.

\textbf{LLM dependency.} Because $S_{\text{factuality}}$ is an NLI entailment score against the generated answer, it depends on \emph{generation style}, not only content. Hedged prefixes (e.g., ``\emph{Based on the provided context\dots}'') lower entailment and thus $T$, even for correct answers. This makes $\tau$ and the weights effectively LLM-specific hyperparameters (Section~\ref{sec:grid}).

%% file: sections/04_experimental_design.tex
\section{Experimental Design}
\label{sec:experimental-design}

\textbf{Datasets.} We use two public benchmarks. \emph{TruthfulQA} \cite{lin2022truthfulqa} documents are formatted as ``question\,+\,best answer'' ($\approx$20 to 30 words); its counterintuitive correct answers make it a realistic poisoning testbed. \emph{FEVER} (Fact Extraction and VERification) \cite{thorne2018fever} documents are enriched to ``question\,+\,verdict\,+\,claim'' ($\approx$30 to 40 words); bare claims of 8 to 12 words caused $\sim$70\% of factuality scores to fall back to the inconclusive $0.5$ baseline, motivating the enrichment. For a software-engineering setting we additionally construct a secure-coding knowledge base of 40 rules curated from the OWASP Top~10 \cite{owasp2021} and CWE \cite{mitrecwe}, evaluated in Section~\ref{sec:usecase}. Each rule is a short natural-language recommendation in the same ``question\,+\,best answer'' format as the benchmark corpora, e.g., ``\emph{How can Structured Query Language (SQL) injection be prevented?}'' paired with parameterized-query guidance referencing CWE-89. The rules were written by the authors as paraphrases of the normative guidance in the source documents, cross-checked against them, and cover the major OWASP categories; the full rule set is included in the released artifacts for independent inspection \cite{repoartifacts}.

\textbf{Protocol.} Each run has two parts: 50 clean queries (Part~A) and the same 50 queries against a corpus with 30\% poisoned documents (Part~B), for 100 samples per run. We evaluate four rule-based poisoning strategies: \emph{contradiction} (appends a contradicting statement), \emph{instruction injection} (appends an override directive), \emph{entity swap} (in-place replacement of entities/numbers, with no appended artifacts), and \emph{subtle manipulation} (false qualifiers/hedging). A \emph{mixed} setting assigns strategies randomly across poisoned documents. Part~A contributes 50 clean samples and Part~B contributes 50 samples of which 30\% ($\approx$15) are labelled poisoned, so a 100-sample run contains about 15 poisoned and 85 clean samples. A naive always-trust baseline that labels every sample clean is therefore correct on 85 clean samples, i.e.\ $\approx$85\% accuracy; hence $\Delta$ and F1, not accuracy, are the informative metrics.

\textbf{Models and metrics.} Documents are embedded with the Sentence-Transformers model \texttt{all-MiniLM-L6-v2} \cite{reimers2019sentence} (384-d) or \texttt{snowflake-arctic-embed2} (1024-d), indexed in Facebook AI Similarity Search (FAISS) (cosine, top-$K$). Generation uses \texttt{llama3.3:70b} (primary) or \texttt{qwen3.5:35b}; NLI runs on a Central Processing Unit (CPU). We report accuracy, precision, recall, F1, and \emph{trust score separation} $\Delta = \bar{T}_{\text{clean}} - \bar{T}_{\text{poisoned}}$, a threshold-independent measure of discriminability. We add 95\% confidence intervals (CIs): Wilson intervals for proportions and a percentile bootstrap ($B{=}20\,000$) for F1.

\textbf{Ground truth.} The clean and poisoned corpora are paired by index, so each query maps to a fixed source document. A sample is labelled \emph{poisoned} when its aligned source document was modified by a poisoning strategy in Part~B, and \emph{clean} otherwise. This document-level label fits the paired benchmark but is stricter than a retrieval-grounded label, which would count a sample as poisoned only when a poisoned document enters the top-$K$ set. Consequently, if the poisoned source document does not enter the retrieved context, the sample still counts as a missed detection although the agent never observed it. Reported recall is therefore a conservative lower bound of what the agent actually sees.

\textbf{Reproducibility.} Generation uses temperature $0.7$, \texttt{max\_tokens}$=512$, and provider-default top-$p$; NLI (\texttt{facebook/bart-large-mnli}) runs on CPU; retrieval uses FAISS (cosine) over $512$-token chunks (overlap $50$) at $K{=}5$. Sampling and poison assignment use a fixed seed ($42$), while LLM decoding stays stochastic (quantified by the variance in Section~\ref{sec:stability}). Calibration fits $\tau$ on a held-out clean split (10th percentile) evaluated on disjoint clean/poisoned samples. All LLMs run on the FARMI/Ollama endpoint.

%% file: sections/05_results.tex
\section{Results}
\label{sec:results}

This section presents the experimental results. It evaluates the Evaluation Agent's detection performance, examines the contribution of its components, tests its robustness across different configurations, and assesses its use in a secure-coding RAG setting.

\subsection{Detection Performance and Per-Strategy Hierarchy}
\label{sec:detection}
On the primary TruthfulQA mixed run (Llama~3.3~70B + MiniLM, $K{=}5$) the agent attains 91\% accuracy (95\% CI 83.8 to 95.2\%), 100\% precision (zero false positives), 40\% recall (19.8 to 64.3\%), F1 57.1\% (25.0 to 80.0\%), and $\Delta{=}0.225$, a +7\% accuracy gain over the naive baseline. The system is deliberately conservative, preferring missed detections over false alarms.

Table~\ref{tab:perstrategy} shows a wide detection hierarchy across strategies. Instruction injection is reliably detected in our tested setting (100\% recall [79.6 to 100\%], F1 96.8\%, $\Delta{=}0.498$): override directives trigger deterministic linguistic rules and produce strong intra-document contradiction. Contradiction is moderate (53.3\% recall [30.1 to 75.2\%]). Subtle manipulation is weak (20\% [7.0 to 45.2\%]) and entity swap is undetected (0\% [0 to 20.4\%], $\Delta{=}0.053$): these in-place edits leave no surface artifacts. With only 15 poisoned samples per run, recall CIs are wide; we treat the per-strategy ranking as indicative, though it is consistent across repeats.

\begin{table}[!t]
\caption{Per-strategy detection on TruthfulQA (100 samples each;
Llama~3.3~70B + all-MiniLM-L6-v2, $K{=}5$).}
\label{tab:perstrategy}
\centering \footnotesize
\begin{tabular}{lccccc}
\toprule
Strategy & Acc. & Prec. & Recall & F1 & $\Delta$ \\
\midrule
Instruction injection & 99\% & 93.8\% & 100\%  & 96.8\% & 0.498 \\
Contradiction         & 92\% & 88.9\% & 53.3\% & 66.7\% & 0.311 \\
Subtle manipulation   & 88\% & 100\%  & 20.0\% & 33.3\% & 0.149 \\
Entity swap           & 85\% & n/a    & 0\%    & 0\%    & 0.053 \\
\midrule
Mixed                 & 91\% & 100\%  & 40.0\% & 57.1\% & 0.225 \\
\bottomrule
\end{tabular}
\end{table}

The dampener explains the extremes: poisoned injection contexts reach mean $P_{\text{poison}}\approx0.99$, consistently activating Eq.~(\ref{eq:damp}), whereas entity-swap contexts stay near $0.32$, below the $0.70$ trigger, so trust scores barely move.

\subsection{Component Ablation (RQ2)}
\label{sec:ablation}
To isolate each signal's contribution (RQ2), we recompute the trust verdict on the primary run by re-weighting the stored per-sample factuality ($F$), consistency ($C$), and poison ($P$) scores at $\tau{=}0.5$; no new generation is required. Table~\ref{tab:ablation} shows that NLI factual verification \emph{alone} is a weak poison detector (0\% recall): by design it scores entailment against the generated answer, not adversarial intent, and reserves strong negative signals for the poison pathway (Section~\ref{sec:approach}). Its role is precision and the trust \emph{baseline}: adding consistency and the poison term raises precision to 100\%, and the non-linear dampener lifts recall from 0\% to 40\%, recovering the full operating point (91\% accuracy, 57.1\% F1). The poison detector alone gives the highest standalone recall (53.3\%, F1 64.0\%) but lower precision (80\%); fusing all three trades recall for zero false positives -- factuality and consistency supply precision, the poison pathway recall.

Using the same offline recomputation, we assessed sensitivity to the manually chosen weights: perturbing each of $\alpha$, $\beta$, and $\gamma$ by up to $\pm0.10$ and renormalizing (125 settings) keeps accuracy at 90.0--91.0\%, F1 at 54.5--57.1\%, and precision at 85.7--100\%, with no setting outperforming the defaults. The operating point is thus robust to moderate weight changes; threshold sensitivity is addressed by the ROC analysis and per-LLM calibration (Section~\ref{sec:stability}).
\input{ablation_components.tex}

\subsection{LLM and Embedding Sensitivity}
\label{sec:grid}
Table~\ref{tab:grid} reports a $2\times2$ factorial study ($K{=}3$). Two findings stand out. First, \emph{LLM choice dominates}: both Llama configurations are identical (91\%, 100\% precision) regardless of embedding model, while both Qwen configurations \emph{underperform the naive baseline} by 14 points (71\%) with precision collapsing to 25 to 28\%. The cause is generation style: Qwen's hedged, verbose answers lower NLI entailment, depressing the mean clean trust score to $\approx$0.64 (vs.\ $\approx$0.83 for Llama) and producing 21 to 23 false positives out of 85 clean samples. Second, embedding dimensionality is near-irrelevant for a concise-output LLM. Thus $\tau{=}0.5$ is implicitly calibrated for Llama, motivating per-LLM calibration.

\begin{table}[t]
\caption{$2\times2$ LLM\,$\times$\,embedding grid (TruthfulQA, 100 samples,
mixed, $K{=}3$).}
\label{tab:grid}
\centering
\footnotesize
\resizebox{\columnwidth}{!}{%
\begin{tabular}{llccccc}
\toprule
LLM & Embedding & Acc. & Prec. & Rec. & F1 & $\Delta$ \\
\midrule
Llama 3.3 70B & MiniLM    & 91\% & 100\%  & 40.0\% & 57.1\% & 0.240 \\
Llama 3.3 70B & Snowflake & 91\% & 100\%  & 40.0\% & 57.1\% & 0.188 \\
Qwen 3.5 35B  & MiniLM    & 71\% & 25.0\% & 46.7\% & 32.6\% & 0.161 \\
Qwen 3.5 35B  & Snowflake & 71\% & 28.1\% & 60.0\% & 38.3\% & 0.199 \\
\midrule
\multicolumn{2}{l}{Naive always-trust} & 85\% & n/a & n/a & n/a & n/a \\
\bottomrule
\end{tabular}}
\end{table}

\subsection{Stability: Variance, Threshold-Independence, and Calibration}
\label{sec:stability}
\input{results_tables.tex}

\textbf{Run-to-run variance.} We re-ran each configuration 2 to 5 times with independent generations. Despite LLM sampling stochasticity, detection metrics are highly stable (Table~\ref{tab:variance}): the primary mixed configuration scores $90.6{\pm}0.5\%$ accuracy and $56.1{\pm}1.3\%$ F1 over five repeats, and injection is invariant at $99.0{\pm}0.0\%$. The low variance indicates that the agent's verdicts are driven by the retrieved evidence rather than by the surface wording of any single generation.
\input{run_to_run.tex}

\textbf{Threshold-independent performance.} The fixed $\tau{=}0.5$ operating point understates the Trust Index. Pooling the mixed-strategy runs, it attains a Receiver Operating Characteristic Area Under the Curve (ROC-AUC) of $0.81$ (Llama), $0.79$ (Mistral~7B), and $0.73$ (Qwen). All three sit well above chance, so Qwen's weak accuracy at $\tau{=}0.5$ is a \emph{thresholding} artifact, not an absence of signal.

\textbf{Per-LLM calibration.} We also add a third LLM, Mistral~7B~Instruct, which reaches $87\%$ accuracy at the default $\tau{=}0.5$ on the primary $K{=}5$ MiniLM mixed run, far above Qwen~3.5~35B ($69\%$ in the same setting; Table~\ref{tab:grid} reports $71\%$ for Qwen at $K{=}3$) despite being five times smaller. This indicates the Trust Index's performance is governed by generation \emph{style} (Qwen's hedged phrasing depresses NLI entailment) rather than model scale. Fitting $\tau$ per LLM as in Section~IV and evaluating on disjoint samples (Table~\ref{tab:calib}) restores baseline-competitive accuracy: Qwen rises from $65.5\%$ to $74.5\%$ by cutting false positives. The three LLMs require different optimal thresholds ($\tau^\star{=}0.71$, $0.58$, $0.43$ for Llama, Mistral, and Qwen), confirming that $\tau$ is LLM-specific. Absolute accuracies in Table~\ref{tab:calib} are lower than in Table~\ref{tab:variance} because the calibration split is deliberately poison-enriched; the meaningful comparison is before vs.\ after calibration on the same split.

\subsection{Retrieval Depth and Overhead}
Increasing retrieval depth from $K{=}3$ to $K{=}5$ leaves the mixed-strategy outcome \emph{identical} (91\% accuracy, 100\% precision, 40\% recall, F1 57.1\%; $\Delta$ changes $0.240\!\to\!0.225$, within noise). The undetected contexts are entity-swap and subtle attacks that leave no textual signal at any depth, so additional documents add no poison evidence. Detection is thus \emph{retrieval-depth-invariant}, with the recall ceiling set by attack difficulty; $K{=}3$ achieves the same result at roughly half the NLI cost.

The evaluation adds $\approx$14.7\,s per sample ($\approx$17$\times$ overhead over baseline RAG), dominated by up to 20 CPU NLI passes at $K{=}5$. This suits batch settings rather than interactive use; Graphics Processing Unit (GPU) inference is projected to cut evaluation below 2\,s. The added cost grows linearly with $K$ and is independent of corpus size, since only the retrieved set is analyzed; in latency-sensitive deployments the agent can run as an asynchronous or batch-mode gate (Section~\ref{sec:discussion}).

\subsection{Cross-Dataset Generalization}
On FEVER, the full 100-sample run reaches 73\% accuracy and \emph{underperforms} the 85\% baseline by 12 points, with separation collapsing to $\Delta{=}0.035$. Although enrichment supplies adequate NLI context, the clean/poisoned trust distributions overlap, indicating that the Trust Index does not generalize out-of-the-box and requires dataset-specific calibration. This negative result bounds the external validity of the TruthfulQA findings and motivates per-domain calibration.

\subsection{Use Case: Secure-Coding RAG Assistant in the SDLC}
\label{sec:usecase}
To ground the agent in a software-engineering setting, we apply it to a \emph{secure-coding assistant}: a RAG-based coding/code-review aid that retrieves from a curated knowledge base of 40 secure-coding rules drawn from the OWASP Top~10 and CWE (e.g., parameterized queries for SQL injection, adaptive password hashing, Transport Layer Security (TLS) configuration). A developer query (``How should passwords be stored?'') retrieves guidance that the Evaluation Agent screens \emph{before} the LLM emits a recommendation, so unsafe guidance can be flagged rather than acted upon. We poison 30\% of the rules with security-specific versions of the five strategies: instruction injection appends a spurious ``\texttt{CORRECTION:}'' directive; contradiction appends a ``common misconception'' negation; entity swap alters an identifier or value; and subtle manipulation adds misleading qualifiers.

\input{seccode_results.tex}

Table~\ref{tab:seccode} reports detection by strategy (Llama~3.3~70B, $K{=}5$), and the pattern follows the benchmark findings. Injected malicious directives are detected almost perfectly (F1 92.3\%, $\Delta{=}0.54$): the agent blocks clear attempts to insert unsafe advice before they reach the developer. Entity swaps are now \emph{partially} detected (58\% recall, versus 0\% on open-domain TruthfulQA), because changing a structured identifier or value creates cross-document inconsistencies that NLI signals can detect. However, contradiction and subtle manipulation still evade detection (0\% recall): a weakened recommendation that presents an insecure practice as acceptable leaves no surface artifact and passes review. The agent can thus guard against overt unsafe-guidance injection in the SDLC, while subtle semantic weakening requires external world-knowledge verification (future work).

%% file: ablation_components.tex
\begin{table}[!t]\caption{Component ablation of the Trust Index (RQ2): detection on the primary TruthfulQA mixed run (Llama~3.3~70B + MiniLM, $K{=}5$, 100 samples), recomputed by re-weighting stored per-sample signals at $\tau{=}0.5$.}\label{tab:ablation}\centering\footnotesize
\begin{tabular}{lcccc}\toprule
Configuration & Acc. & Prec. & Rec. & F1 \\\midrule
Poison detector only & 91.0 & 80.0 & 53.3 & 64.0 \\
NLI only & 84.0 & 0.0 & 0.0 & 0.0 \\
NLI + Consistency & 84.0 & 40.0 & 13.3 & 20.0 \\
Full Trust Index & 87.0 & 100.0 & 13.3 & 23.5 \\
\textbf{Full + Dampener} & 91.0 & 100.0 & 40.0 & 57.1 \\
\bottomrule\end{tabular}\end{table}

%% file: results_tables.tex
\begin{table}[!ht]\caption{Per-LLM threshold calibration (TruthfulQA mixed, $K{=}5$). $\tau$ fit on clean scores only; evaluated on held-out clean + poisoned. ROC-AUC is threshold-independent.}\label{tab:calib}\centering\footnotesize
\resizebox{\columnwidth}{!}{%
\begin{tabular}{llccccc}\toprule
LLM & $\tau$ & Acc. & Prec. & Rec. & F1 & ROC-AUC \\\midrule
\multirow{2}{*}{Llama 3.3 70B} & 0.50 & 84.1 & 96.1 & 40.8 & 57.3 & \multirow{2}{*}{0.807} \\
 & 0.71 & 83.0 & 71.9 & 57.5 & 63.9 & \\
\midrule
\multirow{2}{*}{Qwen 3.5 35B} & 0.50 & 65.5 & 38.4 & 53.3 & 44.7 & \multirow{2}{*}{0.728} \\
 & 0.43 & 74.5 & 51.4 & 40.0 & 45.0 & \\
\midrule
\multirow{2}{*}{Mistral 7B Instruct} & 0.50 & 79.3 & 66.7 & 40.0 & 50.0 & \multirow{2}{*}{0.789} \\
 & 0.58 & 74.1 & 50.0 & 53.3 & 51.6 & \\
\bottomrule\end{tabular}}\end{table}

%% file: run_to_run.tex
\begin{table}[!t]\caption{Run-to-run variance: mean$\pm$std over $R$ independent repeats (TruthfulQA, Llama~3.3~70B + MiniLM, $K{=}5$, 100 samples/run).}\label{tab:variance}\centering\footnotesize
\begin{tabular}{lcccc}\toprule
Strategy & $R$ & Acc.\,(\%) & Recall\,(\%) & F1\,(\%) \\\midrule
Instruction injection & 2 & 99.0$\pm$0.0 & 100.0$\pm$0.0 & 96.8$\pm$0.0 \\
Contradiction & 2 & 92.5$\pm$0.5 & 53.3$\pm$0.0 & 68.1$\pm$1.4 \\
Subtle manip. & 2 & 88.0$\pm$0.0 & 20.0$\pm$0.0 & 33.3$\pm$0.0 \\
Entity swap & 2 & 85.0$\pm$0.0 & 3.3$\pm$3.3 & 5.9$\pm$5.9 \\
Mixed & 5 & 90.6$\pm$0.5 & 40.0$\pm$0.0 & 56.1$\pm$1.3 \\
\bottomrule\end{tabular}\end{table}

%% file: seccode_results.tex
\begin{table}[!t]\caption{Secure-coding RAG assistant: poison detection by attack strategy (40 OWASP/CWE secure-coding rules; Llama~3.3~70B + MiniLM, $K{=}5$).}\label{tab:seccode}\footnotesize
\resizebox{\columnwidth}{!}{%
\begin{tabular}{lccccc}\toprule
Strategy & Acc. & Prec. & Rec. & F1 & $\Delta$ \\\midrule
Instruction injection & 97.5 & 85.7 & 100.0 & 92.3 & 0.542 \\
Contradiction & 85.0 & n/a & 0.0 & 0.0 & 0.156 \\
Subtle manipulation & 85.0 & n/a & 0.0 & 0.0 & 0.066 \\
Entity swap & 72.5 & 29.2 & 58.3 & 38.9 & 0.291 \\
Mixed & 86.2 & 100.0 & 8.3 & 15.4 & 0.163 \\
\bottomrule\end{tabular}}\end{table}

%% file: sections/06_discussion.tex
\section{Discussion}
\label{sec:discussion}

\textbf{Answering the research questions.} \emph{RQ1 (detection effectiveness):} overt poisoning is detected reliably in our tested setting (injection F1 96.8\%, mixed accuracy 91\% with 100\% precision), while in-place edits remain near-undetectable; effectiveness is governed by the visibility of attack artifacts, not by tuning (Section~\ref{sec:detection}). \emph{RQ2 (effect of NLI):} NLI alone is a precision instrument, not a detector: the ablation (Section~\ref{sec:ablation}) shows it supplies the trust baseline and zero false positives, while the poison pathway and dampener supply recall. \emph{RQ3 (resilience):} verdicts are stable across runs, retrieval depths, and embeddings, and carry over to the secure-coding setting, but are bounded by generator style and domain shift (FEVER), which calibration mitigates (Sections~\ref{sec:grid} to~\ref{sec:usecase}). Together these quantify how far the Security-Reliability Gap of Section~I can be closed by an online trust layer.

\textbf{An architectural limit, not a tuning failure.} Entity-swap and subtle attacks change facts in place without textual artifacts. No amount of threshold tuning, reweighting, or deeper retrieval recovers them (Sections~\ref{sec:detection},~\ref{sec:grid}); detecting them requires external world knowledge (e.g., a structured knowledge base). The 40\% mixed-strategy recall ceiling is therefore intrinsic to surface-signal detection.

\textbf{False-positive spillover.} A clean query that retrieves a poisoned neighbor receives elevated poison probability -- security-correct (the environment is contaminated) but costly for precision; relevance-weighted aggregation mitigates but does not eliminate it.

\textbf{Deployment in the SDLC.} The agent fits as a workflow gate: it can screen a secure-coding knowledge base before indexing and, at $\approx$17\,s per query, act as an asynchronous check in code review or continuous integration rather than in the edit loop, with a human reviewing flagged retrievals.

\textbf{Scope and validity considerations.} The poisoning strategies are rule-based and cover common attack patterns, but future work should also evaluate stronger optimization-crafted poisoned passages, including collision-style documents \cite{zou2024poisoned}\cite{zhong2023poisoning}. This study measures \emph{detection} of poisoned context before generation, while measuring whether the LLM adopts injected misinformation, i.e., attack success rate, requires end-to-end evaluation. Relatedly, the document-level ground truth counts a sample as missed even when the poisoned document never enters the top-$K$ context (Section~\ref{sec:experimental-design}), so reported recall is a conservative lower bound. The experiments use controlled sample sizes, with 15 poisoned samples per run, and the secure-coding study focuses on one curated domain. Because NLI scores depend on generation style, per-LLM calibration requires a small labeled clean set for each model.

%% file: sections/07_conclusion.tex
\section{Conclusion and Future Work}
\label{sec:conclusion}
We presented an Evaluation Agent that adds a trust layer to RAG by combining NLI-based factual verification, multi-signal poison detection, and a Trust Index with a non-linear dampener. The agent detects clear poisoning, including 100\% recall for instruction injection and 100\% precision on TruthfulQA, and the secure-coding use case shows it can help block injected unsafe advice in the SDLC. The results also show important boundaries: in-place edits are harder to detect, NLI-based trust scoring depends on the generating LLM, and the FEVER result shows that dataset-specific calibration is important for generalization. Future work includes evaluating stronger poisoning attacks, calibrating the Trust Index across LLMs, adding world-knowledge sources such as, Wikidata, to address entity-swap attacks, accelerating evaluation with GPUs, and measuring misinformation \emph{adoption}, or attack success rate, beyond detection. Three further directions follow from the reviewers' suggestions. First, end-to-end attack evaluation should measure whether poisoned guidance changes generated artifacts, e.g., whether a corrupted secure-coding rule yields vulnerable code or altered developer decisions; this requires code-level security oracles, such as static analyzers, and user studies. Second, evaluation on larger real-world RAG applications with heterogeneous, continuously updated corpora, where calibration data are scarce. Third, jointly optimizing the weights, dampener parameters, and threshold per deployment, extending the perturbation analysis of Section~\ref{sec:ablation}. The implementation, attack generator, and experimental artifacts are available for reproducibility \cite{repoartifacts}.